\documentclass[%
 amsmath,amssymb,
 preprint,%
]{revtex4-1}

\usepackage{graphicx}% Include figure files
\usepackage{dcolumn}% Align table columns on decimal point
\usepackage{bm}% bold math
\usepackage{amsmath}%math
\usepackage{commath}%math formatting
\usepackage{subcaption}%subfigures
\usepackage[dvipsnames]{xcolor}
\usepackage{placeins} % FloatBarrier

\usepackage[utf8]{inputenc}
\usepackage[T1]{fontenc}
\usepackage{mathptmx}
\usepackage{etoolbox}
\usepackage{bbold}
\usepackage{hyperref}

\usepackage[mathcal]{euscript}
\hypersetup{
    colorlinks=true,
    linkcolor=blue,
    filecolor=magenta,      
    urlcolor=cyan,
    }

\usepackage{ulem}

\newcommand{\MnBi}{Mn$_{\textrm{Bi}}$}
\newcommand{\BiMn}{Bi$_{\textrm{Mn}}$}
\newcommand{\Mref}{M^{\mathrm{ref}}}
\newcommand{\muref}{\mu^{\mathrm{ref}}_{\mathrm{eff}}}
\newcommand{\mueff}{\mu_{\mathrm{eff}}}
\newcommand{\alphref}{\alpha^{\mathrm{ref}}}
\newcommand{\MIF}{M_{\mathrm{IF}}}
\newcommand{\MND}{M_{\mathrm{ND}}}
\newcommand{\fMnBi}{f_{\mathrm{Mn_{Bi}}}}
\newcommand{\fBiMn}{f_{\mathrm{Bi_{Mn}}}}

\begin{document}

\title{Magnetic measures of purity for MnBi$_2$Te$_4$}% 
\author{M. Chandler Bennett,$^{1, \dag}$ Jeonghwan Ahn,$^{1}$ Anh Pham,$^{2}$ Guangming Wang,$^{3}$ Panchapakesan Ganesh,$^{2,\ddag}$ Jaron T. Krogel$^{1, \S}$}

\address{$^1$Materials Science and Technology Division, Oak Ridge National Laboratory, Oak Ridge, TN 37831, USA}
\address{$^2$Center for Nanophase Materials Sciences, Oak Ridge National Laboratory, Oak Ridge, Tennessee 37831, USA}
\address{$^3$North Carolina State University, Raleigh, North Carolina, 27695, USA}

\date{\today}

\begin{abstract}
    The intrinsically anti-ferromagnetic topological insulator, MnBi$_2$Te$_4$ (MBT), has garnered significant attention recently.  
    The excitement for this layered van der Waals bonded compound stems from its potential to host numerous exotic topological quantum states.  
    For instance, quantum anomalous Hall states are predicted for odd-layer compounds and axion insulator states for the even-layer compounds.
    Unfortunately, the realization of these phenomena has been hindered by experimental challenges such as the existence of negative charge carriers, i.e., electron doping, which have been linked to anti-site defects among the Mn and Bi sub-lattices.  
    Based on high level diffusion Monte Carlo (DMC) and DMC-tuned DFT+U calculations, we provide benchmark quality results for the bulk Mn magnetization as well as for \MnBi{} and \BiMn{} defects.  
    We use this information to refine and extend models that estimate the anti-site defect concentration in actual MBT samples when combined with data from magnetic susceptibility and intermediate field magnetization measurements. 
    Our models are validated through favorable comparison with prior experimental studies that obtained both magnetic and site occupancy data. 
    We then extend our estimates to a larger set of prior samples to identify a probable zone of low defect density that has yet to be reached in synthesis.
    We anticipate our theoretically based magnetic purity measures may be used as minimization targets in the cycle of refinement needed to synthesize MBT samples with low anti-site defect concentrations and more reproducible topological properties.  
\end{abstract}

\maketitle

$^\dag$ bennettcc@ornl.gov
$^\ddag$ ganeshp@ornl.gov
$^{\S}$ krogeljt@ornl.gov

\section{Introduction}
\label{sec:introduction}

\begin{figure}[htbp!]
\centering
\includegraphics[width=0.55\linewidth]{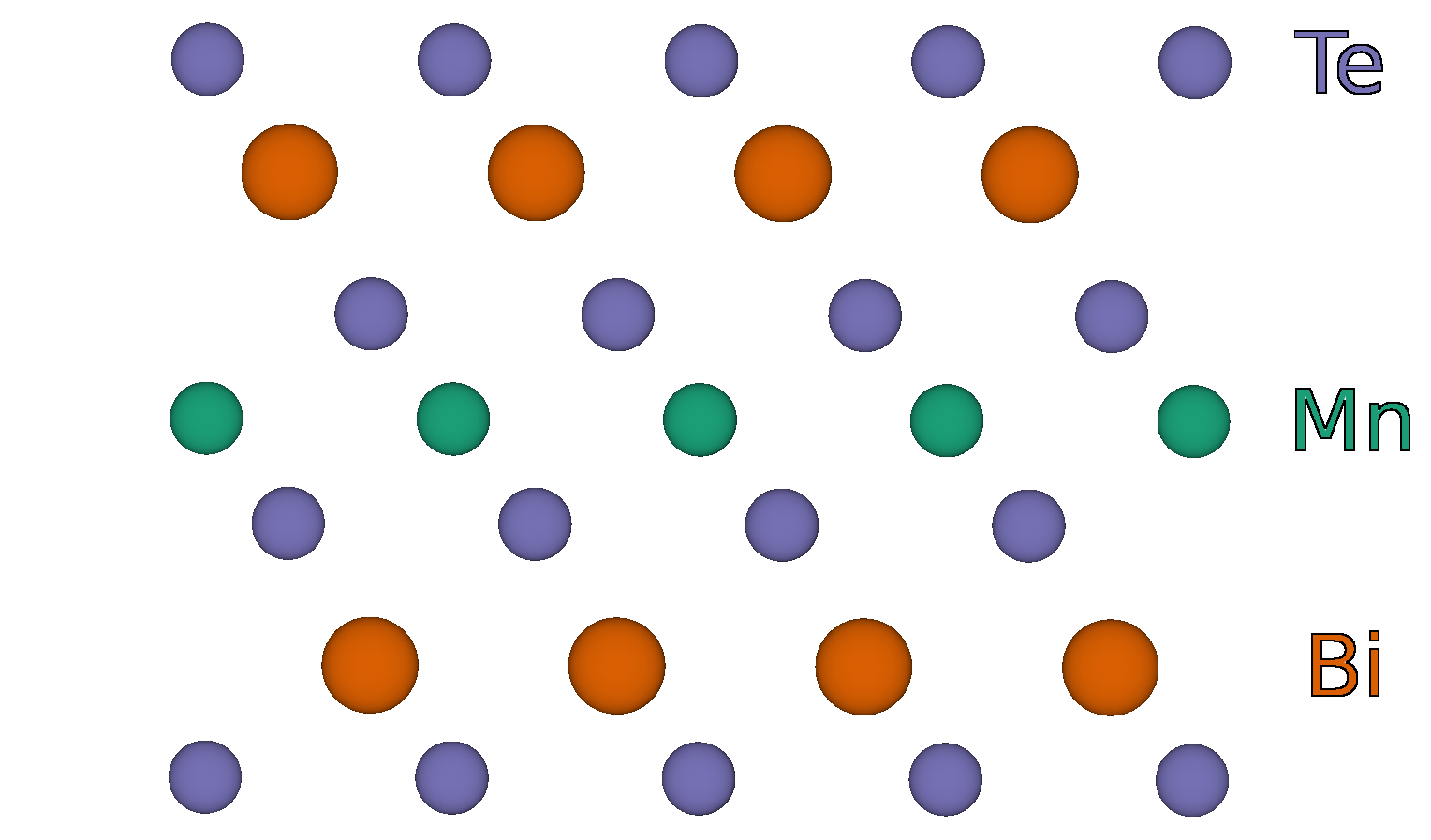}
\caption{
Atomic structure of MnBi$_2$Te$_4$ illustrated by a single septuple layer.  MBT is characterized by a central MnTe layer flanked by two BiTe layers with van der Waals gaps between the septuple layers.
}
\label{fig:crystal}
\end{figure}

The stoichiometric MnBi$_2$Te$_4$ (MBT) crystal is a layered van der Waals bonded anti-ferromagnet.
The material has received considerable attention recently\cite{Otrokov2019, Hao_PhysRevX_2019, Zeugner_Chem_Mater_2019,  yan_crystal_2019, Yan_PhysRevB_2019,  Li_2019_intrinsic, zhang_topological_axion_states_2019, otrokov_unique_thickness_dependent_2019, deng_quantum_2020, liu_robust_2020, Ko_2020, Ding_PhysRevB_2020, Li_PCCP_2020, Lai_PhysRevB_2021, zhao_evenodd_2021, Wang_Innovation_2021} as it is the first observed intrinsically anti-ferromagnetic (AFM) topological insulator (TI) \cite{Otrokov2019}.
As a result, the material provides a promising route to realize exotic topological and quantum phenomena such as the quantum anomalous Hall (QAH) effect, axion insulator states, and Weyl semimetal states, which could be useful within various application domains such as low-energy electronics and topological quantum computation\cite{zhao_evenodd_2021}.
The intrinsic magnetism of MBT also enables the study of magnetic and topological properties together at higher temperatures than previous magnetically-{\it doped} topological insulators\cite{Wang_Innovation_2021,Chang_2013,Chang_2015,Mogi_2015,Checkelsky_2014,Kou_2015}. 
While $T_c$ is $\sim$ 30K in magnetically doped TI compounds, the QAH effect is realized at a very low temperature due to the random distribution of magnetic dopants acting as a source of electron scattering in the edge states even below the Curie temperature. 
As such it is imperative to synthesize MBT crystals/films with minimal magnetic defects to minimize disorder scattering.

MBT consists of Te-Bi-Te-Mn-Te-Bi-Te septuple layers (SLs) as shown in Fig. \ref{fig:crystal}.
Within each SL, the magnetic moments couple ferromagnetically with an easy axis (the $z$-axis) that is perpendicular to the plane \cite{Hao_PhysRevX_2019,yan_crystal_2019,Otrokov2019}.
The moments of neighboring SLs, however, couple anti-ferromagnetically and therefore the material exhibits AFM-$z$ order.
Given that MBT has AFM-z order in the ground state, it preserves ${\bf \mathcal{T}\tau}$ symmetry where ${\bf \mathcal{T}}$ is time-reversal (spin flip) and ${\bf\tau}$ is a fractional lattice translation (here, a 1/2 lattice translation along $z$).
Given this symmetry, MBT can therefore be classified as having $Z_2$ topology with calculated band structures predicting that bulk MBT is an AFM TI \cite{Li_2019_intrinsic,zhang_topological_axion_states_2019,otrokov_unique_thickness_dependent_2019}.
Since few-layer MBT maintains AFM order across layers, by controlling the number of layers in MBT there is potential to arrive at a QAH insulator (odd number of SL layers) or an axion insulator (even number of SL layers).
This ability to switch between the axion state and the QAH state has been experimentally observed in few-layer MBT by switching between even and odd SL layers, respectively \cite{deng_quantum_2020,liu_robust_2020,Ko_2020,Lupke_2022}.

There are, however, a number of challenges related to MBT fulfilling its full promise as an intrinsic magnetic topological insulator. 
MBT has been observed to host $n$-type carriers which can hinder the realization of a high-temperature QAH effect\cite{Wang_Innovation_2021} by pushing the Fermi level into the conduction band which has been observed in transport measurements\cite{Wu_2019_natural, Li_2019_intrinsic, Lee_PhysRevResearch.1.012011, yan_crystal_2019, Zeugner_Chem_Mater_2019}, quasiparticle interference (QPI) measurements using STM\cite{Ko_2020} and ARPES\cite{Wu_2019_natural,Otrokov2019, chen_2019_topological, Hao_PhysRevX_2019, Lee_PhysRevResearch.1.012011, Zeugner_Chem_Mater_2019}. 
Interestingly, QPI measurements show that replacing Bi even with the isoelectronic Sb shifts the Fermi-level down from the conduction band to the center of the bandgap. Given that Sb and Bi have a significant strain mismatch, this suggests that there are intrinsic defects in MBT, and they change due to a changing defect equilibria when Bi is replaced by Sb, thereby controlling the concentration of $\sl n$-type carriers. 
Indeed, strong candidates for the source of the $n$-type carriers are intrinsic defects, with  
recent theoretical evidence~\cite{du_tuning_2021} pointing to the role of anti-site defects, in particular, in providing the unwanted carriers.
It has also been reported that Mn deficiencies (non-stoichiometry) in MBT could be as high as 15\% \cite{yan_A-type_2020}, which can also potentially impact the Fermi level, and magnetic and topological properties.
It is therefore imperative to quantitatively understand the relationship between anti-site defects, the production of $n$-type carriers\cite{li_two-dimensional_2020,Li_PCCP_2020} and resulting bulk magnetism\cite{yan_crystal_2019,Lee_PhysRevResearch.1.012011}.
In this regard, it is highly desirable to synthesize high-quality defect free samples of MBT, that may support the consistent realization of the theoretically predicted QAH phase in the ultra-pure limit.
To that end, having a theoretically derived metric to assess the quality of the sample, as determined by the as-grown anti-site fraction,  
could guide the synthesis of high quality samples and provide a means to directly study how the anti-site fraction is 
coupled to the concentration of $n$-type carriers and the realized topological properties.

Recently, it was observed that high magnetic fields ($\approx$60T) are required to approach saturation of the magnetization exhibited by a pure FM state in MBT\cite{Lai_PhysRevB_2021}.
Moreover, the magnetization profile of MBT samples exhibit stages or multiple plateaus with increasing external field strength, attributed to the presence of \MnBi{} and \BiMn{} antisite defects, which may give the impression that the magnetization is saturated at lower fields ($\approx$8-20T) with reduced moments.
Therefore, magnetic measurements that have not reached full magnetic saturation, and are perceived to be on pristine MBT samples, can lead to underestimates of the Mn local moments. Provided that the moment of \MnBi{} anti-aligns with its neighboring locally ferromagnetic Mn plane, Lai et al.\cite{Lai_PhysRevB_2021} showed that the observed intermediate and high field plateaus in the magnetization could be used to estimate the fraction of anti-site defects when coupled with knowledge of the true Mn local moment in a pure MBT crystal.

In this work, we establish benchmark quality theoretical results for the local magnetization in pristine and defective MBT and use this information to refine and generalize magnetic estimation models for the MBT anti-site defect concentration, similar to the one introduced by Lai et al.\cite{Lai_PhysRevB_2021}.  In particular, we demonstrate that widely available magnetic susceptibility measurements may be effectively combined with magnetization vs. H results, without requiring measurements at very high fields, to constrain or fully determine the relative fraction of \MnBi{} and \BiMn{} defects present in a given sample with support of the theoretical benchmark data.

The theoretical magnetic moments are based on many body diffusion Monte Carlo\cite{grimm1971,foulkes2001} (DMC) calculations and DMC-optimized density functional theory\cite{hohenberg1964,kohn1965} (DFT) calculations.  
Diffusion Monte Carlo is known to provide high quality results for ground state properties of strongly correlated materials\cite{wagner2016_review,kolorenc2010,mitas2010,schiller2015,dzubak2017,saritas2018_lmo,saritas2019_mn_phosphor,rodrigues2020,munoz2020_arxiv,wines2022_accepted,kolorenc2008,foyevtsova2014,wagner2014,zheng2015,yu2015,mitra2015,wagner2015,busemeyer2016,trail2017,saritas2017,yu2017,santana2017_2,sharma2018,saritas2018_coo,kylanpaa2019,wrobel2019,busemeyer2019,ichibha2019,ganesh2020,saritas2020,santana2020,lu2020,huang2021,staros2022}, which are challenging for mean-field approaches.  
For this reason, DMC is often used to benchmark the quality of new density functionals\cite{sun2015} in condensed matter applications\cite{sun2016}.  
The DMC method projects out the lowest energy state that is consistent with the nodes or phase of a given trial wavefunction\cite{anderson1975,anderson1976,ortiz1993}.  
As a rigorous check of the method, we make use of the variational property of the method to optimize the nodal structure of three separate trial wavefunctions within the Hubbard U\cite{anisimov1991} and exact exchange\cite{becke1988} corrected DFT ansatzes.  
Following the independent nodal optimizations, the ground state properties are found to agree closely within each trial wavefunction ansatz, supporting the benchmark quality of the obtained magnetic properties.

In the sections that follow, we develop the magnetic properties and validate key assumptions needed to construct magnetic models of purity for MBT.  
First, we establish a benchmark reference value for the magnetic moment of Mn in pristine MBT with DMC.  
Next, we utilize a DMC-optimized density functional to study the effect of the change in local coordination experienced by \MnBi{} and the Mn sites neighboring \BiMn{} in large supercells.  
Using this optimized functional, we also confirm that the sign of the exchange interaction between \MnBi{} and the Mn-planes is such that the \MnBi{} moment anti-aligns with respect to the locally ferromagnetic ordered moments of each Mn plane, leading to a reduction in the average ground state moment at zero applied field.  
Finally we combine these results to construct additive models linking experimentally measured magnetic properties with the microscopic concentrations of the \MnBi{} and \BiMn{} anti-site species.  
Based on these relations, we estimate the quality of several MBT samples reported in the literature and show that, where information is available, the estimated anti-site defect fractions compare favorably with those obtained by other experimental probes.
These observations support the use of these models to estimate anti-site fractions in synthesized MBT samples with widely available magnetic measurement techniques.  
Since these defects drive the intrinsic $n$-doping in MBT samples, more accessible knowledge of their microscopic concentrations  may enable the community to realize reproducible synthesis routes that tune the Fermi level within the bulk gap to achieve robust QAHE in MBT.

\section{Methods}
\label{sec:methods}

The accuracy of the magnetic moments obtained here derives from the DMC method.  We use DMC as it is implemented in the QMCPACK\cite{kim2018,kent2020} code.  
DMC is a constrained imaginary time projection method. 
Given an approximate solution $\Psi_T$ to the many body time-independent Schr\"{o}dinger equation, referred to as a trial wavefunction, DMC projects out the lowest energy state consistent with the constraint: 
\begin{align}
    \Psi_{DMC} = \lim_{t\rightarrow\infty} e^{-t(\hat{H}-E_{DMC})}\Psi_T 
\end{align}
where $\hat{H}$ is the Hamiltonian.
To ensure that the projected state remains anti-symmetric, the fixed node\cite{anderson1975,anderson1976} (or fixed phase\cite{ortiz1993}) constraint is imposed to restrict the nodal/phase structure of the projected state to match the trial state.
The trial wavefunction employed here is of the Slater-Jastrow\cite{slater1929,condon1930,jastrow1955} type, where the Slater determinant is constructed from the occupied orbitals of DFT calculations using LDA$+U$, PBE$+U$, and PBE0($\omega$)\cite{adamo1999} functionals.  
All DFT calculations are performed with the Quantum Espresso\cite{gianozzi2009} package.
The Jastrow factor included 1, 2, and 3-body correlation terms and was optimized separately for each type of DFT Slater determinant.
Jastrow optimization was performed using the linear method\cite{umrigar2007}, resulting in a variance to energy ratio of 0.012(1) Ha, indicating good quality.
For reasons of computational efficiency, pseudopotentials were used to represent the atomic cores.
Here, we use potentials generated (Mn\cite{Annaberdiyev2018}) or tested (Bi\cite{metz2000}, Te\cite{hermann2002}) within the ccECP\cite{Bennett2017,Bennett2018,Annaberdiyev2018,Wang2019} approach to retain a high degree of accuracy.
The non-local nature of the potentials in the imaginary time DMC projection was handled using the T-moves\cite{casula2005,casula2006} approach and an imaginary timestep of 0.01/Ha was found to be sufficient to converge the magnetic properties of interest.

We calculate the local magnetization at the Mn sites from DMC and DFT for FM and A-AFM magnetic states of MBT by integrating the local contributions about each site.
First, the spin-up and spin-down densities ($\rho^{\uparrow}(\mathbf{r})$ and $\rho^{\downarrow}(\mathbf{r})$, respectively) are computed, and then the spin density is constructed as $\rho^M(\mathbf{r})$, using $\rho^s(\mathbf{r}) = \rho^{\uparrow}(\mathbf{r})-\rho^{\downarrow}(\mathbf{r})$.  
The spin density is normalized to the total number of up electrons minus the total number of down electrons contained in the simulation cell. The estimate of the Mn local magnetization, $M$, is calculated via
\begin{align}
    \begin{split} 
        M_I(R) &= \int_0^R\dif r\int_0^{\pi}\dif\theta\int_0^{2\pi}\dif\phi \,\rho^s(\mathbf{r}-\mathbf{r}_I) \\
        &\equiv \int_0^R\dif r\, M^{\rm AVG} (r)\,,
    \end{split} 
\end{align}
where $I$ indexes an atomic site.  
The integration radius $R$ is extended until the local magnetization stabilizes to its final value. 
In DMC, extrapolated estimators were used to obtain corrected moments. 
While DMC does not explicitly include spin-orbit interactions, the impact of spin orbit on the Mn moments is confirmed to be less than $0.25\%$ by optimized DFT calculations, consistent with the essentially collinear Mn moments observed experimentally in the A-AFM ground state of MBT.

Execution of the simulation workflows and postprocessing of the magnetic results was performed with the Nexus workflow automation package\cite{krogel2016}.  All of the processes and results of this study are made freely available via the Materials Data Facility\cite{Blaiszik2016,Blaiszik2019} [link to be inserted here upon acceptance].

\section{Results and Discussion}
\label{sec:results_and_discussion}

\subsection{Benchmark Magnetic Moment of Mn in Pristine MnBi$_{2}$Te$_{4}$}

First, we seek to establish a benchmark reference value for the magnetic moment of Mn in pristine MBT with the DMC approach. 
The key to our approach is to seek for low energy solutions within sets of trial wavefunctions that efficiently represent the essential physics of the on-site electron-electron interactions within the Mn $d$ manifold.  
The fixed-node variational principle of DMC allows us to identify the most physical projected states as those with the lowest energy.  
Within the Hubbard U framework, we consider trial wavefunctions based on LDA+U and PBE+U Kohn-Sham orbitals and allow U to enter as a variational parameter for direct optimization with DMC.  The U values we consider fall within the range of 0.1-9.0 eV. 
We additionally consider trial wavefunctions based on orbitals obtained from the PBE0($\omega$) hybrid functional. 
In this case, the variational space is extended to include different possible realizations of screening, as represented by the 
exact exchange fraction $\omega$.  
The range of $\omega$ we consider includes variations between 10-40\%.  The three trial wavefunction ansatzes based on LDA+U, PBE+U, and PBE0($\omega$) serve as independent probes of the electronic structure of MBT, where we aim to establish the correct ground state physics.

\begin{figure}[htbp!]
  \centering
  \includegraphics[width=0.7\linewidth]{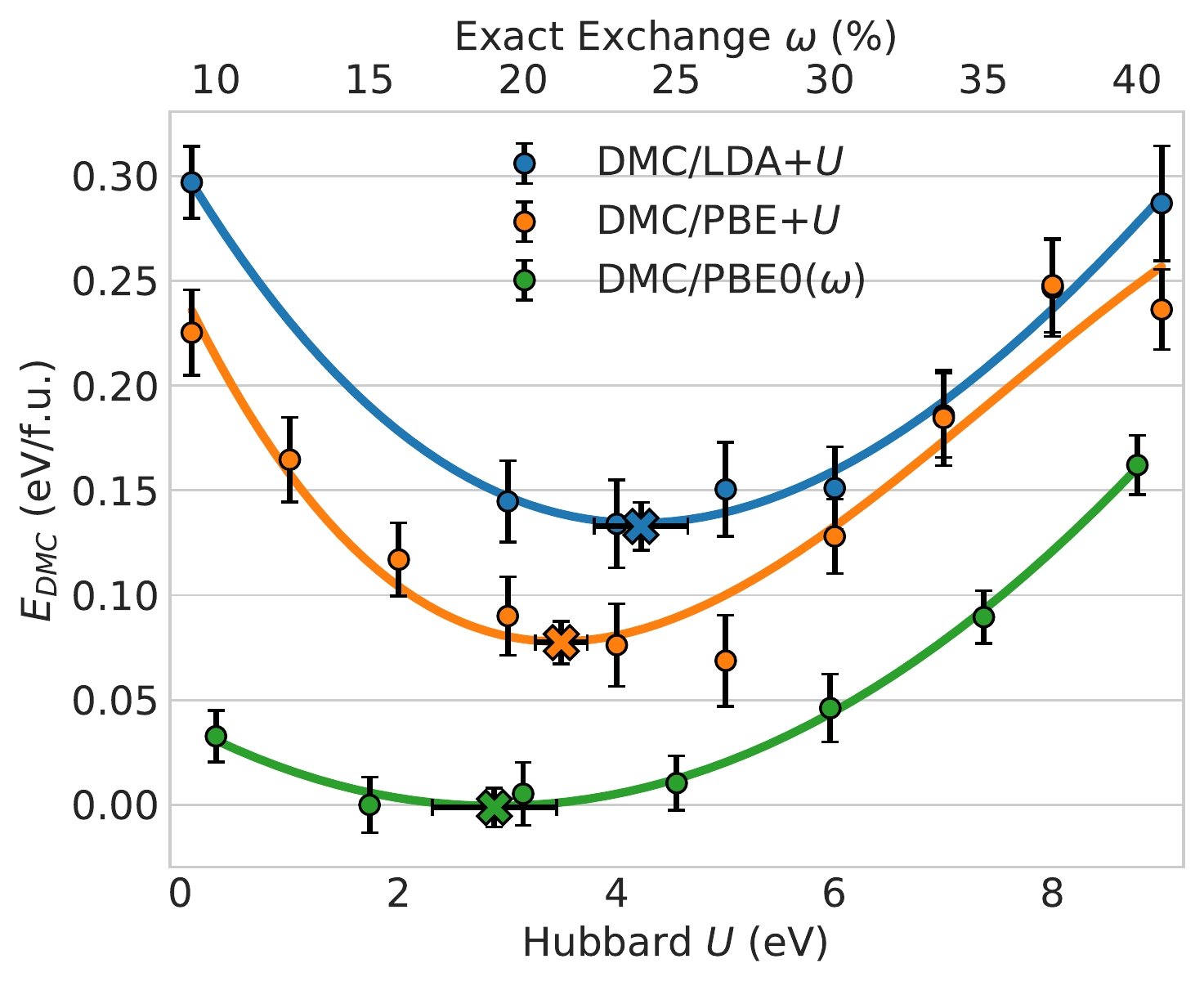}
  \caption{
  DMC nodal optimization using Slater-Jastrow trial functions based on LDA+$U$ (blue), PBE+$U$ (orange), and PBE0($\omega$) (green).  
  DMC total energies per f.u. are shown as colored circles.  The solid lines are cubic polynomial fits to the data and the optimal trial wavefunction parameterizations are marked by the solid X's. The most accurate physical properties are given by DMC/PBE0($\omega$), which has the lowest energy minimum (zero of energy is set to this minimum).
  }
  \label{fig:nodal-opt}
\end{figure}

\begin{table}[!htbp]
  \centering
  \begin{tabular}{l c c c} 
    \hline\hline
    Method    & Optimal Parameter      & M ($\mu_B$)       &  $\mueff{}$ ($\mu_B$) \\
    \hline
    DMC/LDA$+U$        &   U$_{\mathrm{DMC}}=4.4(3)$ eV       &  4.85(2) & 5.76(2)   \\ 
    DMC/PBE$+U$        &   U$_{\mathrm{DMC}}=3.5(2)$ eV       &  4.86(2) & 5.77(2)   \\ 
    DMC/PBE0($\omega$) &  $\omega_{\mathrm{DMC}}=19(2)$ \%    &  4.81(2) & 5.72(2)   \\ 
    \hline
  \end{tabular}
  \caption{
  Onsite Mn magnetizations ($M$) and effective magnetic moments ($\mueff{}$) in MBT calculated by nodal optimized DMC, showing close agreement between the independent \textit{ab initio} predictions.  
  The listed magnetic values include spin-orbit corrections computed via DFT, reducing $M$ and $\mueff{}$ by 0.01 $\mu_B$.   
  The DMC optimized Hubbard $U$ and hybrid exact exchange fraction $\omega$ are also shown along with the corresponding statistical uncertainty. 
  }
  \label{tab:opt-nodal-parameters}
\end{table}

The DMC total energy per formula unit obtained during each of the three nodal optimization processes is shown in Fig. \ref{fig:nodal-opt}. 
In each case, the ferromagnetic state was used for the optimization, which is justified because of the small energy difference between the 
FM and A-AFM MBT states.  
Since this energy difference is on the scale of meV per f.u., it falls well below the accessible statistical errorbars on the total energy and thus the statistics remain the largest source of uncertainty in the optimization.
For DMC/LDA$+U$, we performed the separate nodal optimization with both 7-atom and 28-atom supercells.  
We did not see a statistically significant change to the parameter values for the 28-atom nodal optimizations and thus the primitive cell was used in all subsequent nodal optimizations for pristine bulk MBT.
The minimum energy states were obtained by cubic polynomial fits to the data, as marked on the figure.  
In each case, the statistical uncertainty of the data have been propagated through to the estimated optima via resampling techniques. 
From the results, we see that PBE0($\omega=19\%$) leads to the globally best nodal structure for DMC.
Moreover, the minima of the DMC nodal optimizations based on LDA$+U$ and PBE$+U$ are close by energetically, only differing by 0.14(1) eV and 0.07(1) eV, respectively, from the lowest energy state identified with PBE0-based trial wavefunctions. 

The proximity of the minima in terms of energy also extends to the magnetic properties derived from each optimized ground state wavefunction.
Table \ref{tab:opt-nodal-parameters} summarizes the the optimal $U$ and $\omega$ parameter values found by DMC optimization as well as the corresponding onsite Mn magnetization $M$ and effective magnetic moment $\mueff{}$, which are relevant to magnetization and magnetic susceptibility measurements, respectively. 
The effective atomic magnetic moment for Mn in its high spin $d^5$ configuration is well represented by the spin only contribution, giving $\mueff{}=\sqrt{n(n+2)}~\mu_B$ where $n$ is the number of unpaired electrons\cite{figgis1960}.
Each of the three independently optimized states have very similar Mn magnetization, agreeing to within 0.05(3) $\mu_B$, validating the accuracy of the result. 
Since MBT samples generally show considerable n-type doping \cite{Wu_2019_natural,Otrokov2019, Li_2019_intrinsic, chen_2019_topological, Hao_PhysRevX_2019, Lee_PhysRevResearch.1.012011,  yan_crystal_2019, Zeugner_Chem_Mater_2019}, we have explicitly studied the effect of heavy n-doping on the Mn moments at the optimized DFT level and found it to be negligible (see Supplemental Fig. S2), consistent with the understanding that the lowlying conduction band states in undoped MBT are mainly comprised of Bi and Te $p$-orbitals\cite{trang_crossover_2021}.

The overall lowest energy state, found via DMC/PBE0($\omega$), provides the benchmark magnetic properties for pristine MBT.  
The final reference values are $\Mref{}=4.81(2)~\mu_B$ for Mn magnetization and $\muref{}=5.72(2)~\mu_B$ for the effective moment, which agree well with the refined magnetization of $4.7(1)~\mu_B$ found recently by a combined neutron diffraction/XRD study\cite{Ding_PhysRevB_2020} that was able to explicitly account for the effects of non-stoichiometry and inter-layer mixing.
These reference values will be used as the basis for anti-site fraction estimation models later on, but first we must consider the impact of the local environment on the magnetism of individual \MnBi{} and \BiMn{} defects.

\subsection{Magnetic Properties of \MnBi{} and \BiMn{} Anti-site Defects}

The Mn atoms involved in \MnBi{} and \BiMn{} antisite defects experience altered exchange couplings relative pristine bulk MBT, which may be expected to also influence the local moments and hence the experimentally measured magnetic properties of grown MBT samples.  
In order to reach the system sizes needed to represent these defects, we turn to DFT calculations that have been optimized to retain the accuracy of DMC. 
The DMC fixed node optimization process produces parameterizations of DFT functionals that yield accurate magnetic moments\cite{Kylanpaa_2017}, as we now show for MBT. 

\begin{figure}[htbp!]
  \centering
  \includegraphics[width=0.7\linewidth]{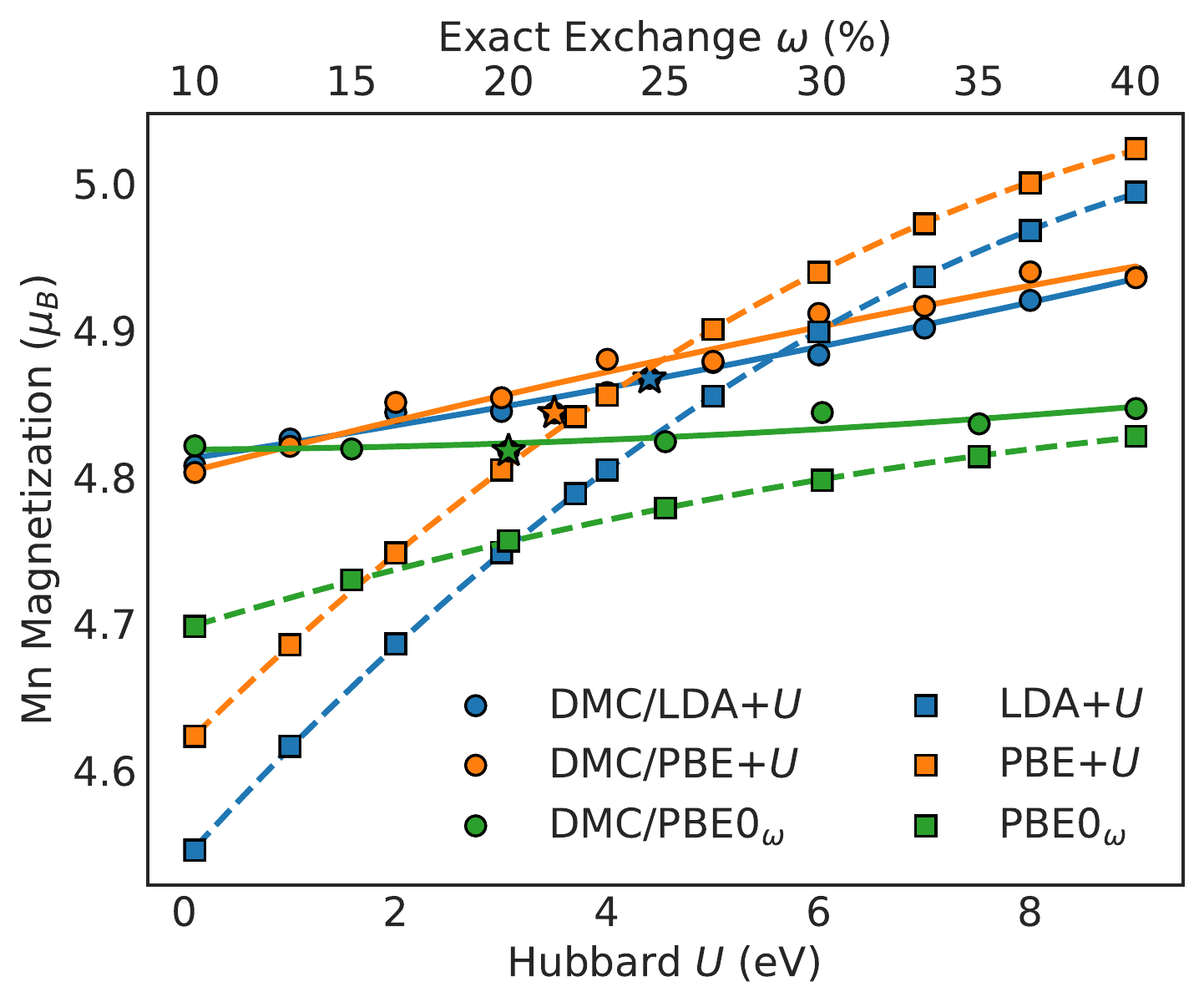}
  \caption{
  DMC (circles, solid lines) and DFT (squares, dashed lines) onsite Mn magnetization as Hubbard $U$ and exact exchange fraction $\omega$ are varied, 
  demonstrating close agreement between the best DMC results (shown as stars) and values calculated with DFT 
  when the optimal DMC $U$ or $\omega$ parameters are used.
  }
  \label{fig:dmc_and_dft_moments_varied}
\end{figure}

In Fig. \ref{fig:dmc_and_dft_moments_varied} we compare the onsite magnetization of Mn in bulk MBT between the LDA+$U$, PBE+$U$, and PBE0($\omega$) density functionals and DMC as the Hubbard U and exact exchange $\omega$ parameters are varied.
The DMC Mn magnetizations clearly show an insensitivity to the nodal structure, with DMC/PBE0($\omega$) showing no statistically significant change, and DMC/LDA$+U$ and DMC/PBE$+U$ changing by only $0.1$-$0.15\mu_B$.
Moreover, the respective DFT results show a higher sensitivity to the parameter values, with a changes close to $0.5 \mu_B$ for LDA$+U$ and PBE$+U$, and a change of $0.1\mu_B$ for PBE0($\omega$).
A key feature of the DFT onsite magnetizations, however, is that despite large sensitivity to the parameter values, they agree well at the DMC optimized parameter values with LDA$+U_{\rm DMC}$, PBE$+U_{\rm DMC}$, and PBE0($\omega_{\rm DMC}$) differing by no more than $0.05\mu_B$ from the reference DMC value.
This supports our approach to use a DMC-optimized DFT functional to predict magnetic properties of the MBT system.
In particular, the DMC optimized PBE+$U$ functional finds a Mn magnetization of $4.83\mu_B$ in near quantitative agreement with the DMC reference value of  $4.81(2)\mu_B$.
For this reason, we select PBE+$U$ with $U_{\textrm{DMC}}=3.5$ eV to study the magnetism of \MnBi{} and \BiMn{} antisite defects.

The isolated \MnBi{} and \BiMn{} antisite defects were represented in large 384 atom supercells using the atomic structures of Ref. \onlinecite{du_tuning_2021}.  
In all calculations, the the magnetic moments of the main Mn-layers were set in the A-AFM arrangement.
For \MnBi{}, we performed calculations with the moment of the off layer Mn atom initialized both aligned and anti-aligned with the nearest, locally FM ordered, Mn plane.  
In each case, the electronic structure relaxed to a ground state with anti-aligned moment, as is assumed experimentally based on the observed behavior of the total magnetization vs. external magnetic field\cite{Lai_PhysRevB_2021}.
The onsite magnetization of \MnBi{} is found to decrease by $0.05\mu_B$, or about 1\%, relative to the in-plane Mn sites and we represent the magnetization of \MnBi{} as $(1-\alphref{})\Mref{}$ with $\alphref=0.01$.  
The Mn magnetization is therefore very robust, but it is still bolstered by the presence of neighboring Mn.  
For \BiMn{}, we consider both the ``missing'' magnetization due to the sudden absence of the on-site Mn, as well as any altered magnetization of the neighboring Mn.  
Our calculations show no significant change of the magnetization of nearby Mn sites, and so the quantitative change in magnetization of \BiMn{} relative to the locally FM Mn plane in bulk is $-\Mref{}$.
The calculated highly localized and small changes in the Mn magnetization of the anti-site defects justifies the construction of additive models of defect concentration, even in the high density limit that has been encountered in multiple experimental MBT samples\cite{Zeugner_Chem_Mater_2019,Ding_PhysRevB_2020,Lai_PhysRevB_2021}.

\subsection{Magnetic Measures of \MnBi{} and \BiMn{} Concentration}

The variable purity of synthesized MBT samples suggests that an optimal synthesis process remains to be found. 
Here we provide a set of additive linear models that combine our benchmark Mn magnetization results with quantities obtainable from experimental magnetization vs. $H$ and magnetic susceptibility measurements.
For single measurements, the models provide constraints on the relative fractions of \MnBi{} and \BiMn{} antisite defects.  With an additional measurement both antisite fractions may be estimated simultaneously.  
We use the models to assess the relative purity of MBT samples presented in prior works\cite{Otrokov2019, Hao_PhysRevX_2019, Zeugner_Chem_Mater_2019, Li_PCCP_2020, Lai_PhysRevB_2021, Yan_PhysRevB_2019} and we compare the effectiveness of the model predictions against other experimental determinations\cite{Zeugner_Chem_Mater_2019, Li_PCCP_2020, Lai_PhysRevB_2021, Ding_PhysRevB_2020} of the anti-site fractions where available.
The relative ease with which some magnetic measurements may be made motivates their potential preferred use in processes of successive synthesis, characterization, and refinement of MBT samples toward the anti-site free limit by employing the model results as optimization targets.

In magnetization measurements performed at low temperature as a function of increasing external magnetic field, MBT is known to exhibit a magnetization plateau at intermediate fields of 8-12T, followed by a higher near-saturation plateau beyond 40T\cite{Lai_PhysRevB_2021}.  
Denoting the height of the intermediate field magnetization plateau as $\MIF{}$, the anti-site fractions $\fBiMn{}$ and $\fMnBi{}$ are constrained by the relation
\begin{align}\label{eq:IF_constraint}
    \frac{\MIF{}}{\Mref{}}=1-\fBiMn{}-2\fMnBi{}(1-\alphref{})
\end{align}
which incorporates the theoretical reference values for Mn in bulk ($\Mref{}=4.81(2)~\mu_B$) and the reduction factor $(1-\alphref{})$ for the \MnBi{} magnetization. 
This relation is useful even in isolation, as both $\fBiMn{}$ and $\fMnBi{}$ are simultaneously bounded by a single measurement at intermediate field. 
This behavior stems from the fact that the presence of \MnBi{} and \BiMn{} independently serve to reduce the observed magnetization below the Neel temperature.
Therefore, as $\MIF{}$ is observed to approach the bulk reference value, both populations of antisite defects become small.

Magnetic susceptibility measurements in the paramagnetic regime beyond the Neel temperature can be used to further constrain the estimated antisite defect fractions.  
From these measurements, Curie-Weiss fits may be used to extract the effective paramagnetic moment, $\mueff{}$, of the magnetic species.  
Since the paramagnetic moment contains contributions from all Mn atoms in the sample, we may estimate it as
\begin{align}\label{eq:MS_constraint}
    \frac{\mueff{}}{\muref{}}=1-\fBiMn{}+2\fMnBi{}
\end{align}
Here, proximity between the measured $\mueff{}$ and the reference ($\muref{}=5.72(2)~\mu_B$) indicates a stoichiometric sample, which may be consistent with small or large anti-site defect populations.  
Therefore the greatest benefit is realized upon combining the magnetic susceptibility constraint with those from other magnetic measurements to fully identify both $\fMnBi{}$ and $\fBiMn{}$.  
The relative availability of both magnetization and magnetic susceptibility measurements may favor this combination in practice for the assessment of sample quality. 

\begin{figure}[htbp!]
  \centering
  \includegraphics[width=0.8\linewidth]{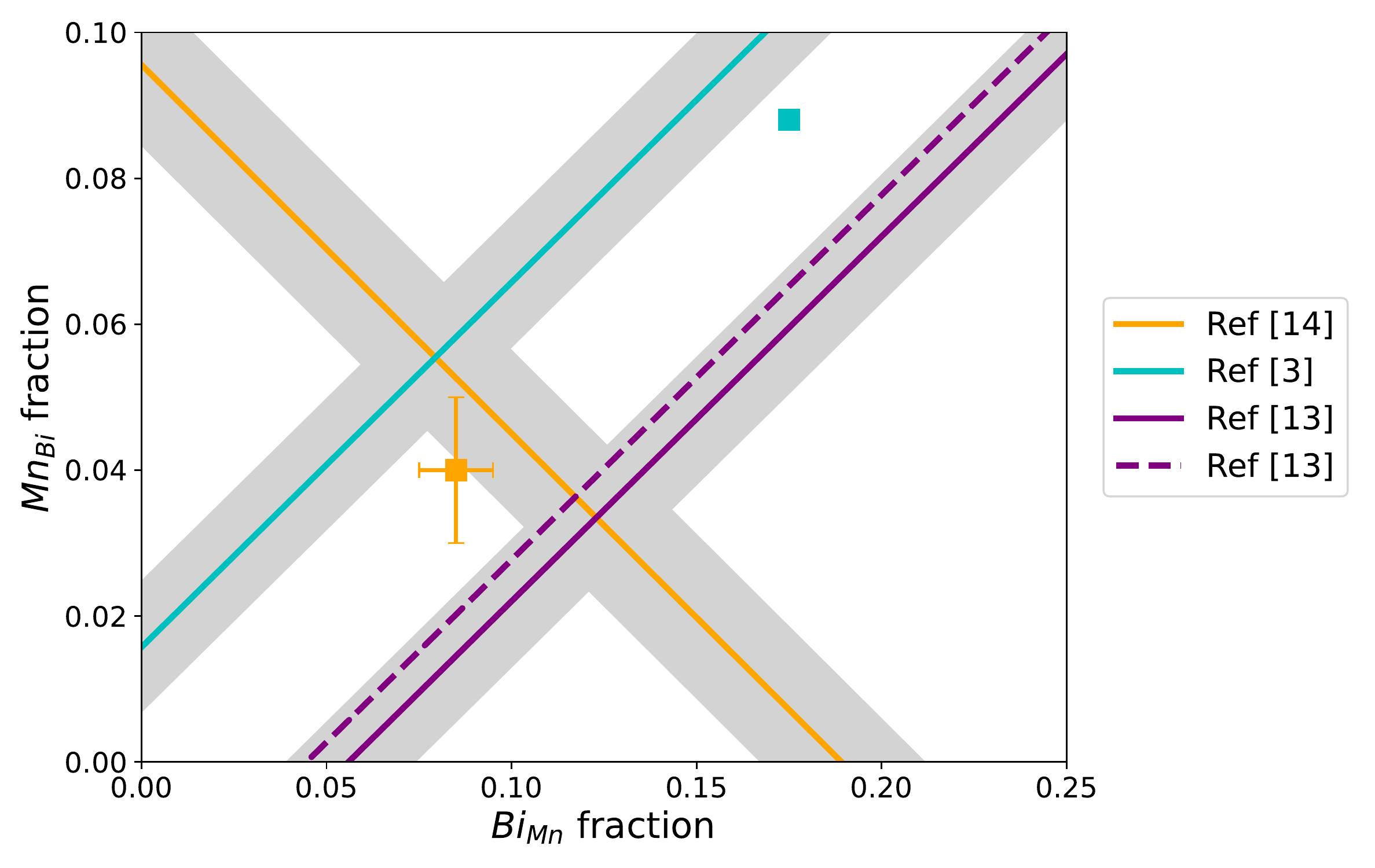}
  \caption{
  Comparison between experimentally determined \MnBi{} and \BiMn{} antisite defect fractions and predictions based on combined benchmark theoretical data and magnetic quantities measured on the same samples\cite{Zeugner_Chem_Mater_2019,Li_PCCP_2020,Lai_PhysRevB_2021}.
  Predicted lines of constraint from Eqs. \ref{eq:IF_constraint} \& \ref{eq:MS_constraint} are shown using magnetic data from Table \ref{table:exp_mag}, with joint theoretical and experimental uncertainty shown in grey.   Experimentally determined \MnBi{} and \BiMn{} fractions\cite{Zeugner_Chem_Mater_2019,Lai_PhysRevB_2021} and stoichiometric Bi:Mn ratio constraint\cite{Li_PCCP_2020} from the same studies are shown as filled squares and a dashed line for comparison with the predictions marked in the same color. The overall agreement in each case serves as an independent check on the predictions.
  }
  \label{fig:exp_mag_metric_comp}
\end{figure}

For validation purposes, in Fig. \ref{fig:exp_mag_metric_comp} we compare experimentally determined anti-site defect fractions with predicted $\fBiMn{}$ and $\fMnBi{}$ lines of constraint obtained from Eqs. \ref{eq:IF_constraint}-\ref{eq:MS_constraint} and experimental data for $\MIF{}$ and $\mueff{}$ from the same studies.  
Shaded regions around the predicted constraint lines include combined experimental and theoretical uncertainties.    
The experimental magnetic data are listed in Table \ref{table:exp_mag}, along with available experimentally determined $\fBiMn{}$ and $\fMnBi{}$ from the same studies.  
In the figure, solid squares mark the experimentally determined concentrations\cite{Zeugner_Chem_Mater_2019,Lai_PhysRevB_2021} and one stoichiometric determination from ICP-MS\cite{Li_PCCP_2020} is also included as a dashed line.  Corresponding experimentally determined and predicted values are shown in the same color for ease of comparison. 

\begin{figure}[htbp!]
  \centering
  \includegraphics[width=0.8\linewidth]{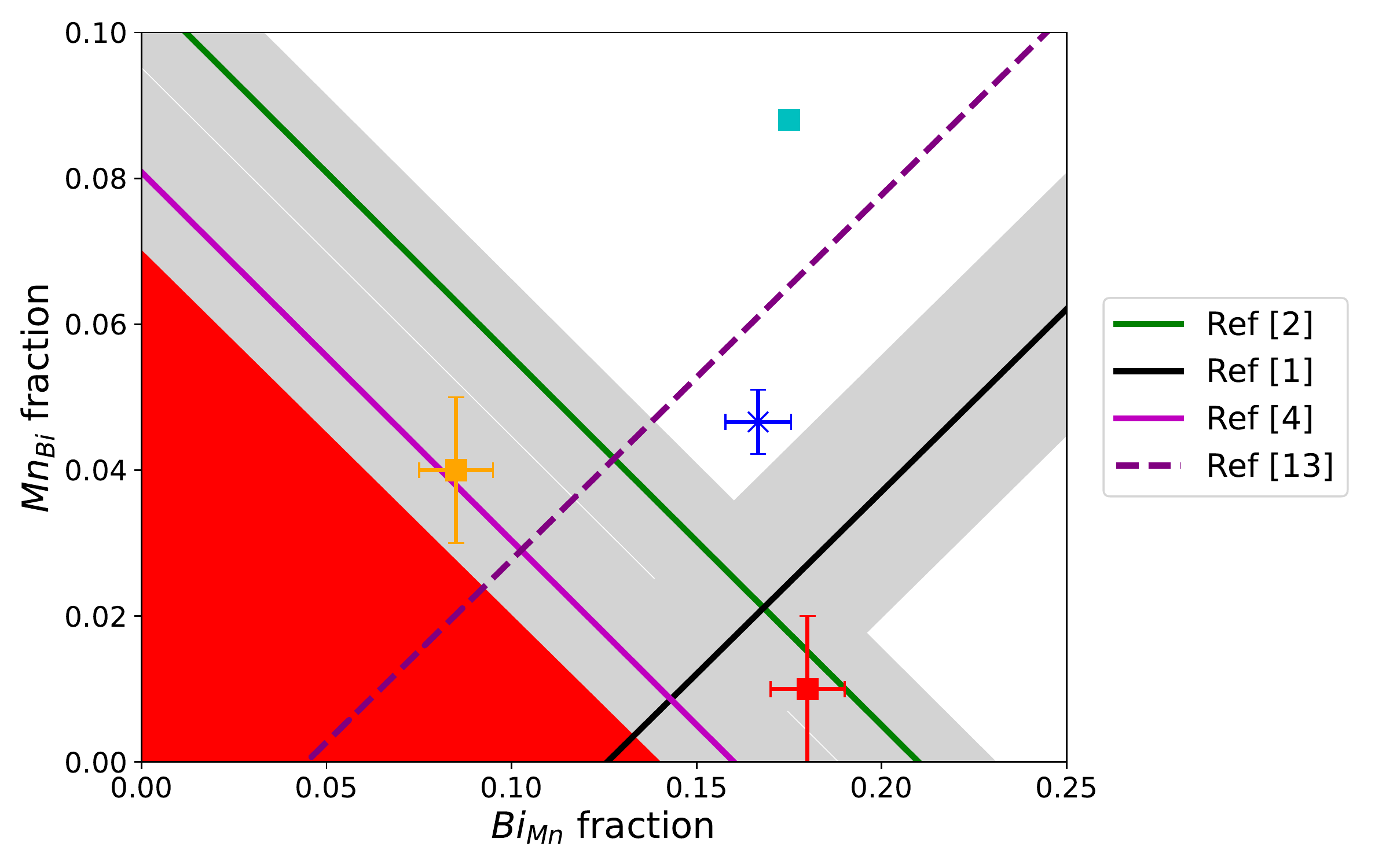}
  \caption{
  Predicted \MnBi{} and \BiMn{} antisite defect fractions of synthesized MBT samples based on experiments\cite{Hao_PhysRevX_2019,Otrokov2019,yan_crystal_2019,Yan_PhysRevB_2019} lacking microscopic information about the defects, shown as lines of constraint when only one magnetic measurement was available, and as crosses when both antisite fractions could be isolated.  Antisite fraction data obtained directly in other experimental studies\cite{Zeugner_Chem_Mater_2019,Lai_PhysRevB_2021,Ding_PhysRevB_2020,Li_PCCP_2020} are shown as solid squares and a dashed line of constraint.  The red region denotes a probable exclusion zone, where samples with lower defect density have yet to be synthesized. 
  }
  \label{fig:exp_mag_metric_unique}
\end{figure}

Overall, the concentrations obtained from magnetic quantities via Eqs. \ref{eq:IF_constraint}-\ref{eq:MS_constraint} agree well with those obtained by other means.  
In Ref. \onlinecite{Zeugner_Chem_Mater_2019}, the antisite fractions were determined to be $\fBiMn{}=17.5(1)\%$ and $\fMnBi{}=8.8(1)\%$ via x-ray site occupancy refinements.  
For the same sample, magnetic susceptibility measurements gave $\mueff{}=5.9(1)~\mu_B$.  
Using Eq. \ref{eq:MS_constraint} and $\fMnBi{}=8.8(1)\%$ we predict a value of $15(2)\%$ for $\fBiMn{}$, which is statistically consistent with the experimental value.  
Considering instead the estimation of $\fMnBi{}$, we obtain $10(1)\%$ versus $8.8(1)\%$ experimentally.  
In Ref. \onlinecite{Lai_PhysRevB_2021}, the intermediate field magnetization plateau was estimated as $\MIF{}=3.9(1)~\mu_B$, while antisite fractions of $\fBiMn{}=8-9\%$ and $\fMnBi{}=4\%$ were estimated by considering the difference between the intermediate and high field (60T) magnetization plateaus.  
Using Eq. \ref{eq:IF_constraint} and either $\fBiMn{}=8.5(1.0)\%$ or $\fMnBi{}=4(1)\%$, we predict $\fMnBi{}=5(1)\%$ or $\fBiMn{}=11(3)\%$, respectively, again agreeing well with the experimental values.
As a final point of comparison, we consider magnetic susceptibility and stoichiometry data from Ref. \onlinecite{Li_PCCP_2020}.  
In that work, the Bi:Mn ratio was determined to be $2.14(1)$ by ICP-MS, while $\mueff{}=5.4$ was obtained from magnetic susceptibility measurements\footnote{We used the $\mueff{}$ reported for the $H^{ab}$ magnetic susceptibility data since these data showed a clear linearity in the inverse susceptibility with respect to temperature, enabling the low field $\mueff{}$ to be estimated with lower bias from associated choices in the fitting range.}. 
Similar to the relations above, the Bi:Mn ratio may be estimated as $(3\muref{}/\mueff{}-1)$, for which we predict a ratio of $2.18(6)$ from our benchmark $\muref{}$.
The observed agreement supports the use of our magnetic estimation measures to assess the quality of MBT samples where the anti-site fractions remain unknown.

We combine previously known anti-site fraction data\cite{Zeugner_Chem_Mater_2019,Lai_PhysRevB_2021,Ding_PhysRevB_2020,Li_PCCP_2020} with predictions based on Eqs. \ref{eq:IF_constraint}-\ref{eq:MS_constraint} for other experimental studies providing only magnetic data\cite{Hao_PhysRevX_2019,Otrokov2019,yan_crystal_2019,Yan_PhysRevB_2019} in Fig. \ref{fig:exp_mag_metric_unique}.   
The combined data represent samples from eight separate studies, providing useful information regarding the purity of existing synthesis attempts.
As before, experimentally determined data are shown in solid squares and the dashed line as in Fig. \ref{fig:exp_mag_metric_comp}, while predicted constraints from magnetic data are shown as solid lines with uncertainty in grey.
One case where $\fMnBi{}$ and $\fBiMn{}$ could be estimated based on data from two types of magnetic measurements\cite{Yan_PhysRevB_2019} is shown with an ``X''.
The predicted antisite fractions have an uncertainty of about 1\%, which, while being primarily driven by the uncertainty in the magnetic measurements, remains similar in magnitude to most of the independent experimental probes.
From the figure, we see that a large \BiMn{} population is encountered in at least half of the samples, with three\cite{Zeugner_Chem_Mater_2019,Yan_PhysRevB_2019,Ding_PhysRevB_2020} samples falling near $\fBiMn{}=17\%$ and another\cite{Otrokov2019} with $\fBiMn{}>15\%$.  
Additionally, 75\%\cite{Otrokov2019,Yan_PhysRevB_2019,Zeugner_Chem_Mater_2019,Lai_PhysRevB_2021,Ding_PhysRevB_2020,Li_PCCP_2020} of the samples clearly obey $\fBiMn{}\ge 2\fMnBi{}$, while half of the banded region allowed by the remaining two\cite{Hao_PhysRevX_2019,yan_crystal_2019} is also consistent with this inequality.  
This suggests that MBT samples with a larger number of \MnBi{} than \BiMn{} are uncommon (Mn-rich conditions), with the opposite tendency occurring often. 
This observation is also consistent with the widely observed n-doping in MBT, as   
DFT calculations in Ref. \onlinecite{du_tuning_2021} predicted \BiMn{} to often favor a donor configuration. 
Therefore persistent high \BiMn{} concentrations may play a key role in setting the Fermi level above the bulk band gap.   
Finally, we note that the combined data suggests a probable region that may not have been reached by current synthesis efforts, as shown in red in Fig. \ref{fig:exp_mag_metric_unique}.
This region naturally forms a synthesis target, the realization of which we hope may be facilitated by monitoring and refinement of the growth process based on the MBT purity measures presented in this work.

\section{Conclusions}
\label{sec:conclusions}

We have provided accurate benchmark information regarding the onsite Mn magnetization in pure and defective MnBi$_2$Te$_4$ with high-level diffusion Monte Carlo methods, combined with DMC optimized DFT.
The onsite Mn magnetization in pristine bulk was validated through the close agreement of three independent variational DMC calculations based on trial wavefunctions obtained from optimal parameterizations of Hubbard-U and exact exchange density functionals.
Using a DMC optimized DFT functional, we were able to capture the experimentally assumed anti-alignment of \MnBi{} moments relative to neighboring Mn planes and obtain an accurate description of the impact of \MnBi{} and \BiMn{} anti-site defects on the magnetization of MBT samples. 
Our benchmark value for the onsite Mn magnetization in bulk of $\Mref{}=4.81(2)~\mu_B$ is in good agreement with recently obtained single crystal neutron diffraction estimates of $4.7(1)~\mu_B$\cite{Ding_PhysRevB_2020} on a sample with verified low concentration of \MnBi{} defects. 

We have discussed how our theoretical benchmark results can be combined with accessible experimental magnetic measurements to infer microscopic details of the \MnBi{} and \BiMn{} anti-site defect fractions in a given MBT sample. 
We have further shown that the models developed here reproduce well the antisite defect fractions estimated by other experimental means as a validation. 
A potential advantage offered by these models is that the desired microscopic information may be acquired from more accessible experimental magnetic probes, such as magnetization measurements performed below 12T and conventional paramagnetic susceptibility measurements.
Therefore, we anticipate that these measures of purity of MBT with respect to antisite defects may prove useful to guide the targeted synthesis of MBT samples toward the idealized material.  
This is consequential because as this limit is approached the intrinsic n-type doping of MBT is also reduced, with potentially less severe residual strategies being required to bring the material to the point of bulk neutral occupations, where the desired topological properties are most likely to be realized in a consistent and reproducible manner.

\section*{Acknowledgements}

The authors would like to thank M.-H. Du for sharing structural data regarding large defective MBT supercells\cite{du_tuning_2021}. 
This research has been supported by the US Department of Energy, Office of Science, Basic Energy Sciences, Materials Sciences and Engineering Division, as part of the Computational Materials Sciences Program and Center for Predictive Simulation of Functional Materials. 
This research used resources of the Compute and Data Environment for Science (CADES) at the Oak Ridge National Laboratory, which is supported by the Office of Science of the U.S. Department of Energy under Contract No. DE-AC05-00OR22725. An award of computer time was provided by the Innovative and Novel Computational Impact on Theory and Experiment (INCITE) program.
This research used resources of the Argonne Leadership Computing Facility, which is a DOE Office of Science User Facility supported under Contract DE-AC02-06CH11357.

This manuscript has been authored by
UT-Battelle, LLC under Contract No. DE-AC05-00OR22725 with the U.S. Department of Energy. The United States Government retains and
the publisher, by accepting the article for publication, acknowledges that the United States Government retains a nonexclusive,
paid-up, irrevocable, worldwide license to publish or reproduce the published form of this manuscript, or allow others to do so,
for United States Government purposes. The Department of Energy will provide public access to these results of federally sponsored
research in accordance with the DOE Public Access Plan (http://energy.gov/downloads/doe-public-access-plan).

\begin{table}[hbt!]
 \begin{tabular}{l l l l l l} 
    \hline
    \hline
    Reference                                           & $\MIF{}$ ($\mu_B$)  & $\mueff{}$ ($\mu_B$) &  $\fBiMn{}$ (\%)   &  $\fMnBi{}$ (\%)  \\  [0.5ex] 
    \hline
    Ref. \onlinecite{Otrokov2019} &      & 5.0(2)  &          &         \\
    Ref. \onlinecite{Hao_PhysRevX_2019}                 & 3.8  &         &          &         \\
    Ref. \onlinecite{Zeugner_Chem_Mater_2019}           &      & 5.9(1)  &  17.5(1) &  8.8(1) \\
    Ref. \onlinecite{Yan_PhysRevB_2019}                 & 3.56 & 5.3(1)  &          &         \\
    Ref. \onlinecite{Li_PCCP_2020}                      &      & 5.7     &          &         \\
    Ref. \onlinecite{Lai_PhysRevB_2021}                 & 3.9  &         &  8-9     &  4      \\
    \hline
    \hline
\end{tabular}
\caption{ 
 Experimentally measured MnBi$_2$Te$_4$ intermediate field magnetization plateau ($\MIF{}$) from low temperature $M$ vs. $H$ measurements, in-plane magnetization ($\MND{}$) from neutron diffraction, and effective paramagnetic magnetic moment ($\mueff{}$) from magnetic susceptibility measurements.  Where available, experimental estimates of \BiMn{} and \MnBi{} antisite defect fractions are also shown.  
 }
\label{table:exp_mag}
\end{table}

\clearpage

\providecommand{\noopsort}[1]{}\providecommand{\singleletter}[1]{#1}%

\end{document}